\documentclass[twocolumn,aps,nofootinbib]{revtex4-2}
\usepackage{graphicx}
\usepackage{epstopdf}
\usepackage{latexsym}
\usepackage{amsmath}
\usepackage{amssymb}
\usepackage{color}
\usepackage{mathrsfs}

\usepackage[center]{subfigure}

\begin{document}

%%%%%%%%%%%%%%%%%%%%%%%%%%%%%%%%%%%%%%%%%%%%%%%%%%%%%%%%%%%%%%%
 \newcommand{\bq}{\begin{equation}}
 \newcommand{\eq}{\end{equation}}
 \newcommand{\bqn}{\begin{eqnarray}}
 \newcommand{\eqn}{\end{eqnarray}}
 \newcommand{\nb}{\nonumber}
 \newcommand{\lb}{\label}
\newcommand{\PRL}{Phys. Rev. Lett.}
\newcommand{\PL}{Phys. Lett.}
\newcommand{\PR}{Phys. Rev.}
\newcommand{\CQG}{Class. Quantum Grav.}
 %%%%%%%%%%%%%%%%%%%%%%%%%%%%%%%%%%%%%%%%%%%%%%%%%%%%%%%%%%%%%%%

\title{Hawking radiation received at infinity in higher dimensional Reissner-Nordstr\"om black hole spacetimes}

\author{Kai Lin $^{1, 2}$}
\author{Wei-Liang Qian$^{2,3,4}$}
\author{Xilong Fan$^{5}$}
\author{Bin Wang$^{3}$}
\author{Elcio Abdalla$^{6}$}

\affiliation {1) Hubei Subsurface Multi-scale Imaging Key Laboratory, Institute of Geophysics and Geomatics, China University of Geosciences, Wuhan 430074, Hubei, China}
\affiliation{2) Escola de Engenharia de Lorena, Universidade de S\~ao Paulo, 12602-810, Lorena, SP, Brazil}
\affiliation{3) Center for Gravitation and Cosmology, College of Physical Science and Technology, Yangzhou University, Yangzhou 225009, China}
\affiliation{4) Faculadade de Engenharia de Guaratinguet\'a, Universidade Estadual Paulista, Guaratinguet\'a, SP, Brazil}
\affiliation{5) School of Physics and Technology, Wuhan University, Wuhan 430072, China}
\affiliation{6) Instituto de F\'isica, Universidade de S\~ao Paulo, S\~ao Paulo, Brazil}
%

%\date{\today}
\date{April 20th, 2021}

\begin{abstract}
In this work, we investigate the Hawking radiation in higher dimensional Reissner-Nordstr\"om black holes as received by an observer, resides at infinity.
The frequency-dependent transmission rates, which deform the thermal radiation emitted in the vicinity of the black hole horizon, are evaluated numerically.
Apart from the case of four-dimensional spacetime, the calculations are extended to higher dimensional Reissner-Nordstr\"om metrics, and the results are found to be somewhat sensitive to the spacetime dimension.
In general, it is observed that the transmission coefficients practically vanishes when the frequency of the emitted particle approaches zero.
It increases with increasing frequency and eventually saturates to some value.
For four-dimensional spacetime, the above result is shown to be mostly independent of the metric's parameter, neither of the orbital quantum number of the particle, once the location of the event horizon, $r_h$, and the product of the charges of the black hole and the particle $qQ$ are given.
For higher-dimensional cases, on the other hand, the convergence becomes more slowly.
Moreover, the difference between states with different orbital quantum numbers is found to be more significant.
As the magnitude of the product of charges $qQ$ becomes more significant, the transmission coefficient exceeds one.
In other words, the resultant spectral flux is amplified, which results in an accelerated process of black hole evaporation.
The relation between the calculated outgoing transmission coefficient with existing results on the greybody factor is discussed.

\end{abstract}


\maketitle

\section{Introduction}
\renewcommand{\theequation}{1.\arabic{equation}} \setcounter{equation}{0}

The four laws of black hole mechanics were initially proposed~\cite{agr-bh-thermodynamics-01} as merely an analogy to the four laws of thermodynamics. 
The notion of the Bekenstein-Hawking entropy~\cite{agr-bh-thermodynamics-02,agr-bh-thermodynamics-03,agr-bh-thermodynamics-04} shed light on the microscopic degrees of freedom of the black hole.
It plays an important role in the holographic principle~\cite{adscft-holo-01,adscft-holo-03} and the AdS/CFT correspondence~\cite{adscft-02,adscft-01}.
From a different aspect, Hawking's approach~\cite{agr-hawking-radiation-01,agr-hawking-radiation-02} shows that the black hole emits radiation according to a thermal spectrum, which, in turn, demonstrates the consistency with Bekenstein's results.
Wilczek and coauthors~\cite{agr-hawking-radiation-06,agr-hawking-radiation-07} have further considered the effect of self-interaction correction to the metric.
By employing a semi-classical approximation~\cite{agr-hawking-radiation-07}, the related physical process is interpreted as a particle traversing the horizon from inside while moving inward.
Mathematically, the interpretation is closely related to the fact that the relevant contribution only concerns a small interval bounded by the initial and final radii, in the immediate vicinity of the horizon~\cite{agr-hawking-radiation-07}.
As a result, the calculated tunneling rate is with respect to the position of the horizon.

The above semi-classical method introduced by Wilczek has incited many subsequential studies (see, for instance, Refs.~\cite{agr-hawking-radiation-09,agr-hawking-radiation-10}).
Furthermore, it has inspired other approaches~\cite{agr-hawking-radiation-12,agr-hawking-radiation-13,agr-hawking-radiation-14,agr-hawking-radiation-15}.
Angheben {\it et al.} has proposed~\cite{agr-hawking-radiation-13} to evaluate the imaginary part of the action via the Hamilton-Jacobi equation, which is an extension from the approach by Srinivasan and Padmanabhan~\cite{agr-hawking-radiation-12}.
The method can be applied to static metrics, which might be singular at the horizon. Moreover, the proposed procedure is independent of any particular choice of spatial coordinates.
However, the formal solution of the Hamilton-Jacobi equation partly relies on the symmetries of the specific metric.
Moreover, as the particle's self-gravitation is ignored, the resultant particle emission rate takes into account only the leading term linear in energy. 
The method has been examed in the context of a broader class of spacetimes, as well as different types of fields, where consistent results are obtained~\cite{agr-hawking-radiation-14,agr-hawking-radiation-16,agr-hawking-radiation-17,agr-hawking-radiation-18}.

The particle emission took place at the horizon of a black hole experience an effective potential during its course to the spatial infinity. 
In other words, the resultant spectral flux received by an observable at infinity is further deformed by a frequency-dependent transmission coefficient, $\gamma(\omega)$.
To be specific, for an observer resides at infinity, the expectation value for the number of a given particle species of frequency $\omega$ reads
\bqn
\lb{greybody}
\langle n(\omega)\rangle = \frac{\gamma{\omega}}{e^{\beta\omega}\pm 1}\ ,
\eqn
where $\beta$ is the inverse of Hawking temperature, the plus (minus) sign is for fermions (bosons).
Here 
\bqn
\lb{greybody_factor}
\gamma(\omega)=\left|\frac{\mathcal{T}}{\mathcal{I}}\right|^2
\eqn
is the so-called greybody factor~\cite{agr-bh-superradiance-01}, where $\mathcal{T}$ and $\mathcal{I}$ are the amplitudes of the transmission and incident waves.
In the literature, however, the latter are usually defined in the context of an incoming wave from infinity with given frequency $\omega$, interpreted as the probability for it to reach the horizon of the black hole.
Nonetheless, it can be shown~\cite{agr-bh-superradiance-06} that the above probability coincides with that for an outgoing wave in the mode $\omega$ to escape to infinity through the effective potential of the black hole.
Therefore, regarding Wilczek's viewpoint of Hawking radiation, it measures the tunneling probability for penetration of the barrier governed by the black hole metric.
In asymptotically flat spacetimes, it is directly associated with the S-matrix element.

At small frequency, analytical results on the greybody factor can be obtained by perturbative approach~\cite{agr-bh-superradiance-03}.
On the other hand, for frequencies with large imaginary part, the monodromy method~\cite{agr-qnm-05} has been utilized~\cite{agr-bh-superradiance-04,agr-bh-superradiance-06}.
Estimations on the bound of the greybody factor have also been carried out~\cite{agr-bh-superradiance-07}.
However, in general, as the forms of the effective potentials are rather complicated, the exact solution for a given metric is not straightforward.
As a result, one usually resorts to numerical approaches.

The present study involves an attempt to numerically investigate the Hawking radiation as well as the frequency-dependent transmission coefficient in Reissner-Nordstr\"om black hole spacetime.
The rest of the paper is organized as follows.
In the next section, we briefly review the Hawking radiation on the horizon of the black hole.
The frequency-dependent transmission coefficient is obtained numerically for various types of fields in section III.
Additional discussions and concluding remarks are given in the last section.

\section{Tunneling radiation by the semi-classical approach}
\renewcommand{\theequation}{2.\arabic{equation}} \setcounter{equation}{0}

In this section, we briefly review the Hawking radiation at the horizon of the Reissner-Nordstr\"om black hole, in terms of the Hamilton-Jacobi method~\cite{agr-hawking-radiation-13}.
The background $n$ dimensional metric and electromagnetic potential are given by
\bqn
\lb{metric}
ds^2&=&-f(r)dt^2+\frac{dr^2}{f(r)}+r^2d\Omega_{n-2}\nb\\
dA&=&A_t(r)dt
\eqn
where $f=1-\frac{8\Gamma((n-1)/2)M}{(n-2)\pi^{(n-3)/2}r^{n-3}}+\frac{4\Gamma((n-1)/2)Q^2r^{2(3-n)}}{(n-2)(n-3)\pi^{(n-3)/2}}$ and $A_t=\frac{Qr^{3-n}}{(3-n)}$. $d\Omega_{n-2}$ is a $n-2$ dimensional unit sphere.
$M$ and $Q$ are the mass and charge of the black hole, respectively.
The event horizon $r_h$ and inner horizon $r_i=br_h$ satisfy the relation $M=\frac{(n-2)\pi^{(n-3)/2}(1+b^{n-3})}{8\Gamma((n-1)/2)}r_h^{n-3}$ and $Q=\frac{b^{n-3}(n-2)\pi^{(n-3)/2}}{2\Gamma((n-3)/2)}r_h^{2(n-3)}$.

As a semi-classical approximation, the dynamics of particles with various spin satisfy the Hamilton-Jacobi equation~\cite{agr-hawking-radiation-12,agr-hawking-radiation-13,agr-hawking-radiation-16,agr-hawking-radiation-17,agr-hawking-radiation-18}, namely,
\bqn
\lb{HJ1}
g^{\mu\nu}\left(\frac{\partial S}{\partial x^\mu}-qA_\mu\right)\left(\frac{\partial S}{\partial x^\nu}-qA_\nu\right)+m^2=0
\eqn
where $m$ and $q$ are the mass and charge of particle.
For a static four-dimensional metric, one may look for a solution in the form
\bqn
\lb{HJ1b}
S=-\omega t+R(r)+Y(\theta,\phi,\cdot\cdot\cdot) .
\eqn

By substituting the specific forms of the metric and electromagnetical potential into above equation, one finds the following radial equation after separating the variables
\bqn
\lb{HJ2}
-\frac{1}{f}\left(\omega+qA_t\right)^2+fR'^2+m^2=\frac{\lambda}{r^2}
\eqn
where $\lambda$ is a constant. 
Therefore, near the event horizon, we have
\bqn
\lb{Tunnling1}
R&=&\int dr \frac{\sqrt{(\omega-\omega_h)^2-f(r)(m^2-\lambda/r^2)}}{f(r)} \nb\\
&\to&\int \frac{d\sigma}{\sigma}\frac{2\sqrt{(\omega-\omega_h)^2-f'(r_h)(r-r_h)(m^2-\lambda/r^2)}}{f'(r_h)}  \nb
\eqn
where $\omega_h=-qQ/r_h$. 
$\sigma=\int\frac{dr}{\sqrt{f}}=\frac{2\sqrt{r-r_h}}{\sqrt{f'(r_h)}}$ is the leading contribution of the invariant distance.
The integral is carried out by deforming the contour to avoid the singularity at the horizon, which picks up half a residue:
\bqn
\lb{Tunnling2}
\Im S=\Im R=\frac{2\pi(\omega-\omega_h)}{f'(r_h)} .
\eqn
It is noted that the result is invariant with respect to time recalibration and spatial diffeomorphism~\cite{agr-hawking-radiation-13}.
Subsequently, the quantum tunneling rate from horizon is given by
\bqn
\lb{Tunnling3}
\Gamma=\exp\left(-2\Im S\right)=\exp\left(-4\pi\frac{\omega-\omega_h}{f'(r_h)}\right)
\eqn
and the Hawking temperature at horizon reads
\bqn
\lb{Tunnling4}
T_h=\frac{f'(r_h)}{4\pi}.
\eqn
At this point, the Hawking radiation is purely thermal.

In what follows, we proceed to evaluate the role of the effective potential on the resultant spectral flux, as the emitted particle further penetrates the barrier toward an observer at spatial infinity.

\section{The frequency-dependent transmission coeﬀicient in Reissner-Nordstr\"om spacetime}

\renewcommand{\theequation}{3.\arabic{equation}} \setcounter{equation}{0}

Usually, for a non-rotating metric, the equation of motion of various fields can be simplified by using the method of separation of variables and the radial part of the resultant field equation reads 
 \bqn
\lb{Schrodinger}
\frac{d^2\Psi}{dr_*^2}+\left((\omega+qA_t)^2-V(r)\right)=0 .
\eqn
The above equation is Schrodinger-type, where $r_*=\int dr/f$ is tortoise coordinate and $V$, the effective potential, is governed by the specific spacetime as well as particle state. 
For asymptotically flat spacetimes, $V(r\rightarrow r_h)=0$ and $V(r\rightarrow\infty)=V_\infty$, so the solutions possess the following asymptotic forms at horizon and infinity~\cite{agr-bh-superradiance-01}
\bqn
\lb{Boundary}
\Psi\sim
  \begin{array}{cc}
    \mathcal{R}e^{-ik_Hr_*}+\mathcal{I}e^{ik_Hr_*} & r\rightarrow r_h ,\\
    \mathcal{T}e^{ik_\infty r_*}+\mathcal{A}e^{-ik_\infty r_*} & r\rightarrow \infty .\\
  \end{array}
\eqn
The equation of motion implies that the Wronskians
 \bqn
\lb{Wronskian}
W(r\rightarrow r_h)=-2ik_h(\left|\mathcal{R}\right|^2-\left|\mathcal{I}\right|^2),\nb\\
W(r\rightarrow \infty)=2ik_\infty(\left|\mathcal{T}\right|^2-\left|\mathcal{I}\right|^2).\nb\\
\eqn
are conserved so that one finds the following relation
\bqn
\lb{relation}
\left|\mathcal{T}\right|^2-\left|\mathcal{A}\right|^2=-\frac{k_H}{k_\infty}(\left|\mathcal{R}\right|^2-\left|\mathcal{I}\right|^2).
\eqn
For the present scenario, the incident wave is propagates outward with amplitude $\mathcal{I}$ and one also requires
\bqn
\lb{Boundary2}
\mathcal{A}\equiv0 .
\eqn
In the case of a massless scalar field, the relevant effective potential in Reissner-Nordstr\"om spacetime can be derived from the Klein Gordon equation, which reads
\bqn
V=\frac{f}{r^2}\left[l(l+n-1)+\frac{n-2}{2}rf'+\frac{(n-4)(n-2)}{4}f\right] ,\nb\\
\eqn
where $l$ is the orbital quantum number of the particle state.
%For electromagnetic and gravitational fields with odd parity, the corresponding effective potentials are found to be~\cite{agr-qnm-10,agr-qnm-11,agr-qnm-asympototic-iteration-03}
%\bqn
%V_{\pm}=r^{-2}f\left(l^2+l-q_\pm+4Q^2r^{-2}\right) ,
%\eqn
%with
%\bqn
%q_\pm=\left(3\pm\sqrt{9+16Q^2(l-1)(l+2)}\right)/2 .
%\eqn

At this point, the problem is reduced to that of the one-dimensional barrier penetration. 
To obtain the transmission coefficient, we resort to solve Eq.~\eqref{Schrodinger} numerically with the boundary conditions Eqs.~\eqref{Boundary} and~\eqref{Boundary2}. 

For the present study, we make use of an approach based on numerical integration.
We relegated to the appendix technical details of the numerical scheme, and present the results in Figs.~\ref{Figwb}-\ref{Figql}.

\begin{figure*}[tbp]
\centering
\includegraphics[width=1\columnwidth]{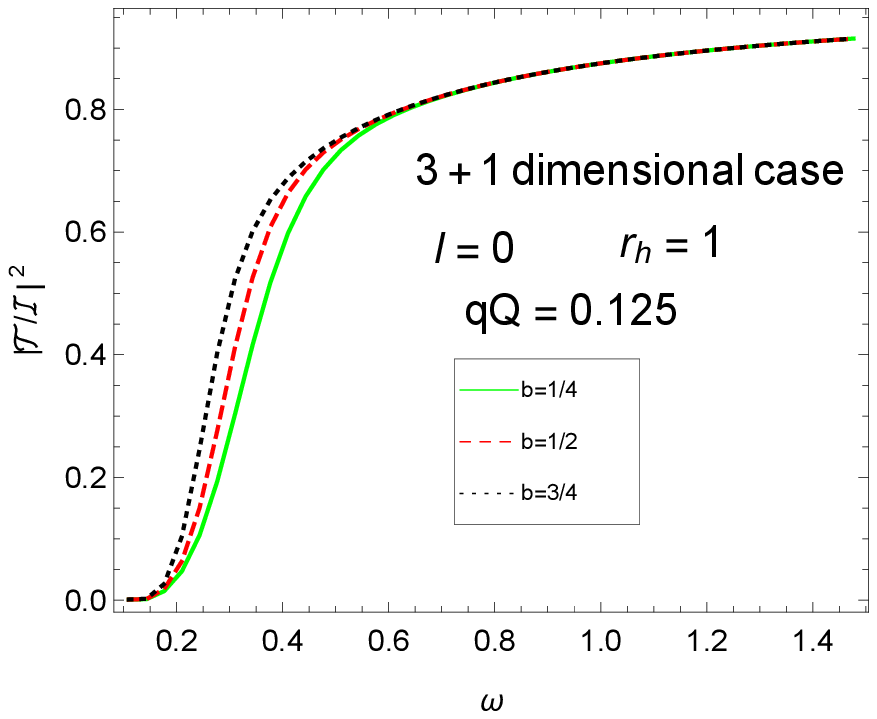}\includegraphics[width=1\columnwidth]{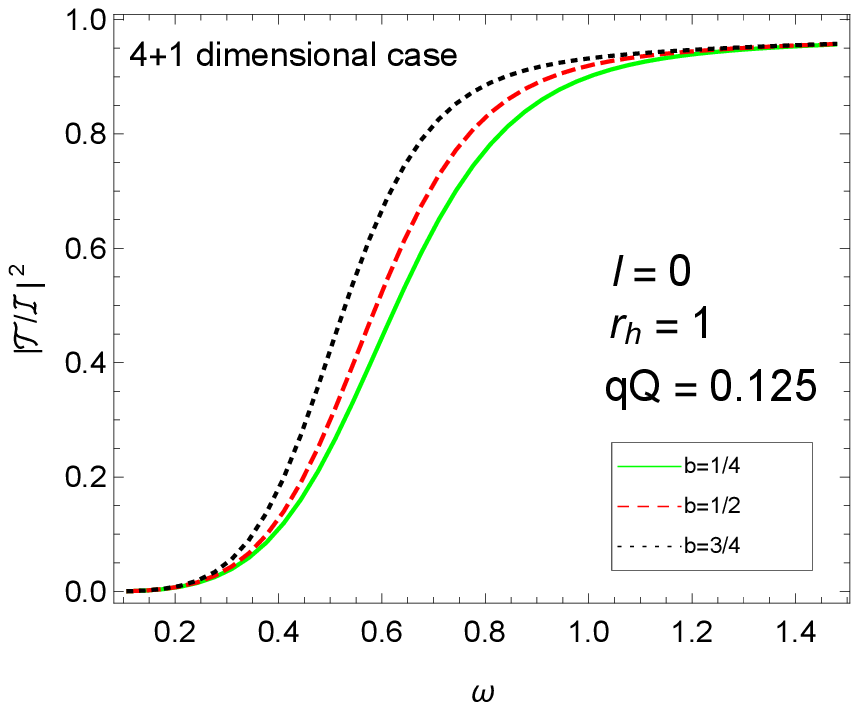}
\includegraphics[width=1\columnwidth]{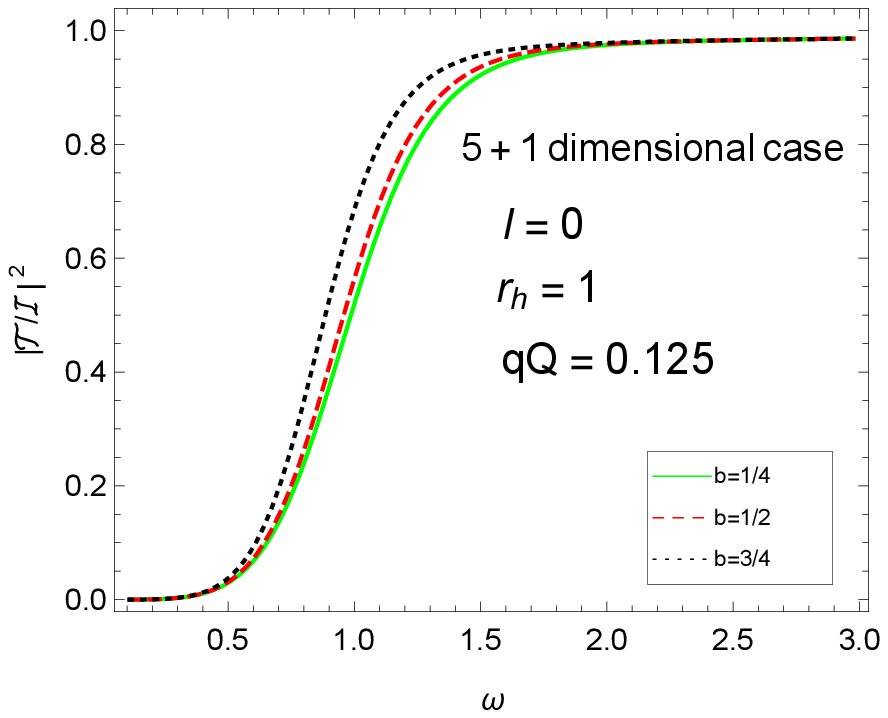}\includegraphics[width=1\columnwidth]{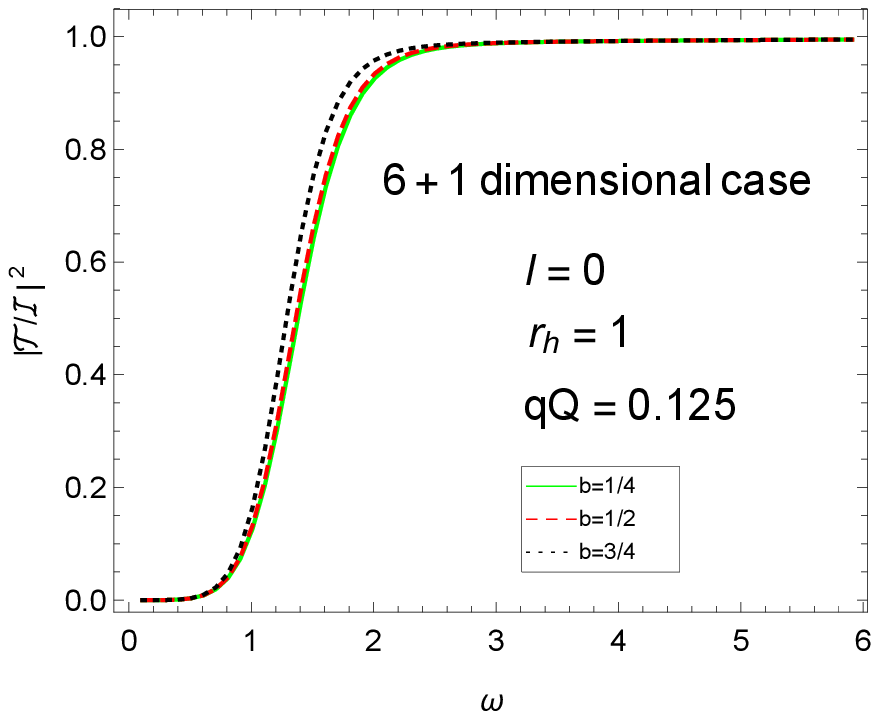}
\caption{
The calculated transmission coefficient $\left|\frac{\mathcal{T}}{\mathcal{I}}\right|^2$ for massless scsalar field as a function of the frequency $\omega$ for different values of $b$.
The calculations are carried out with $r_h=1$, $qQ=\frac{1}{8}$, and $l=0$.}
\lb{Figwb}
\end{figure*}

\begin{figure*}[tbp]
\centering
\includegraphics[width=1\columnwidth]{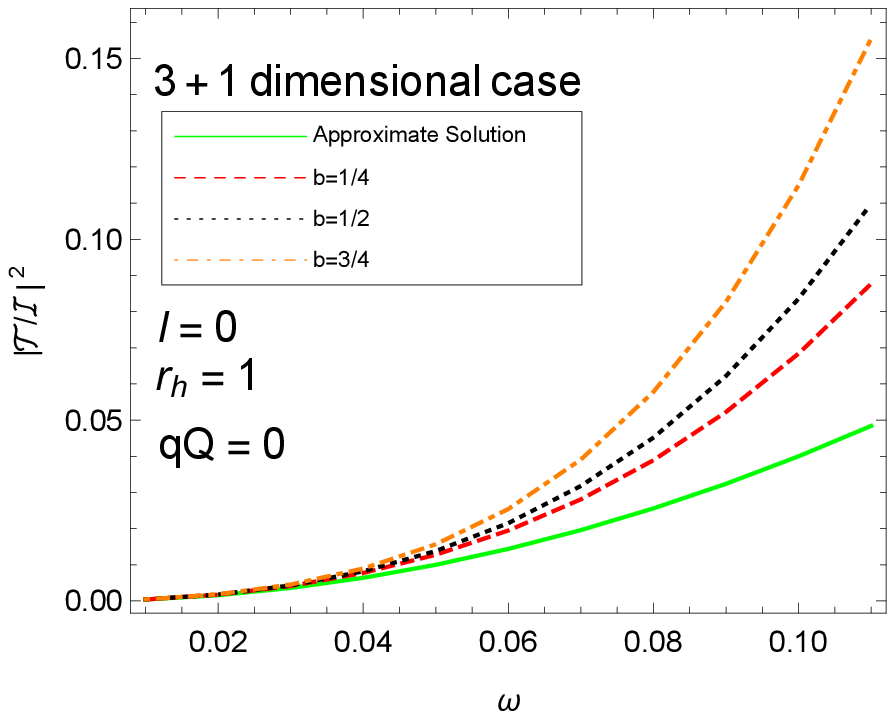}\includegraphics[width=1\columnwidth]{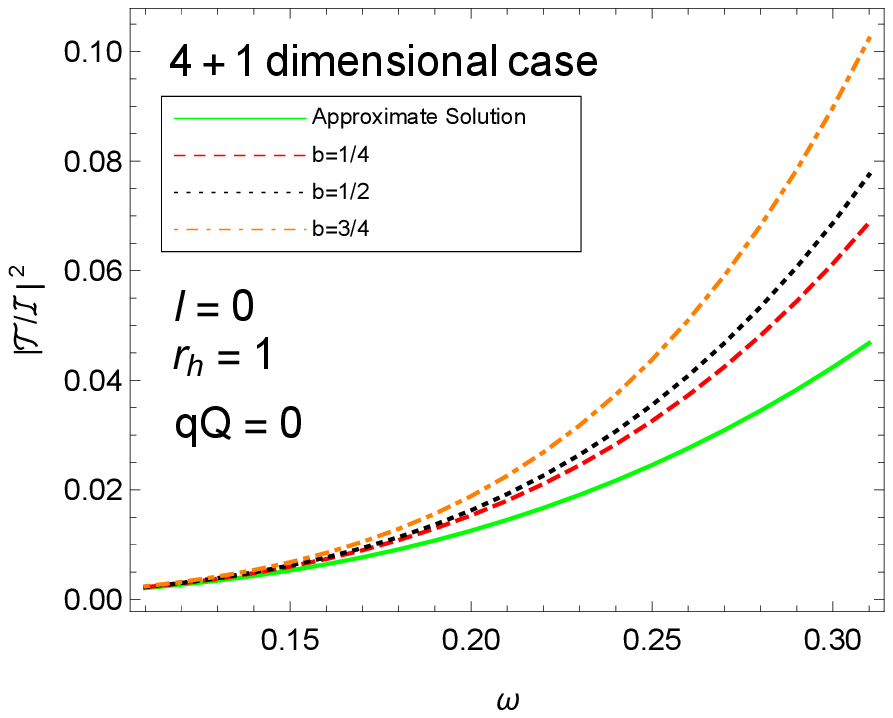}
\includegraphics[width=1\columnwidth]{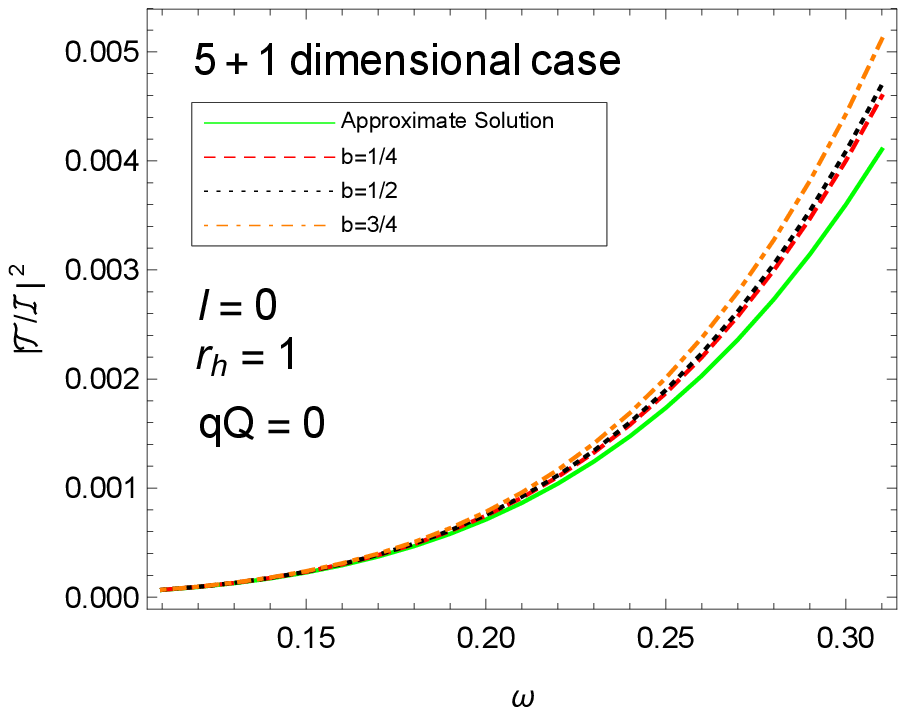}\includegraphics[width=1\columnwidth]{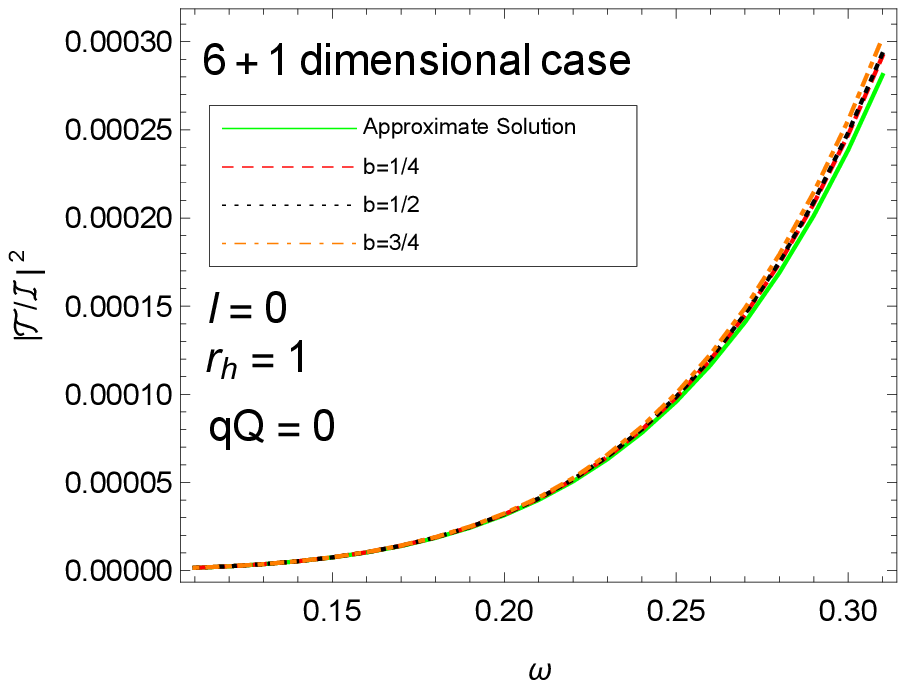}
\caption{
The same as Fig.~\ref{Figwb}, but for the low frequency region $\omega \ll T_h$ and $\omega r_h \ll 1$. 
The calculated results are compared with the low frequency limit obtained in Refs.~\cite{agr-bh-superradiance-05,agr-bh-superradiance-06}.}
\lb{Figweb}
\end{figure*}

\begin{figure*}[tbp]
\centering
\includegraphics[width=1\columnwidth]{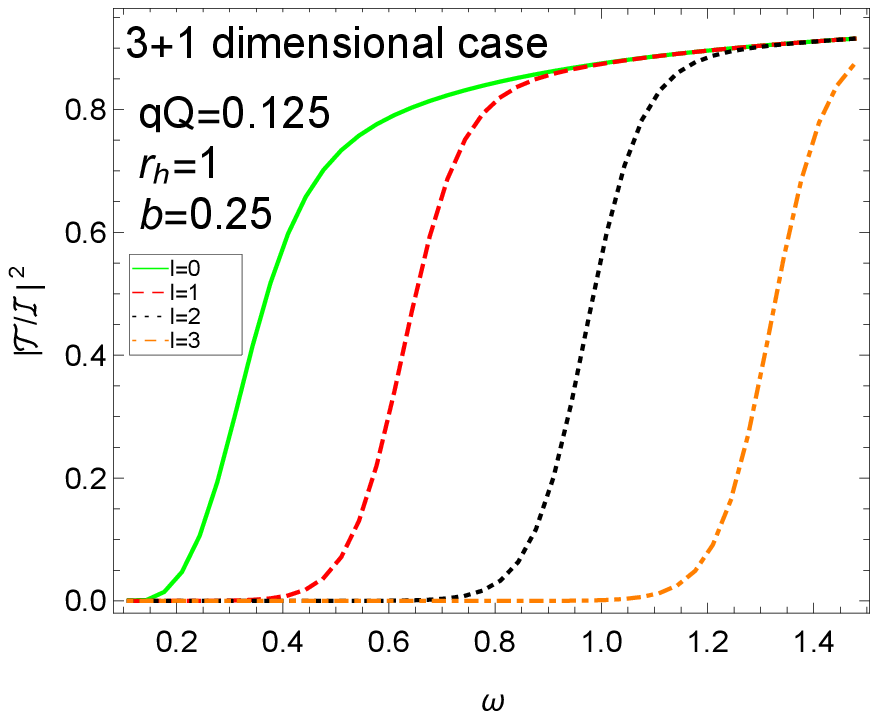}\includegraphics[width=1\columnwidth]{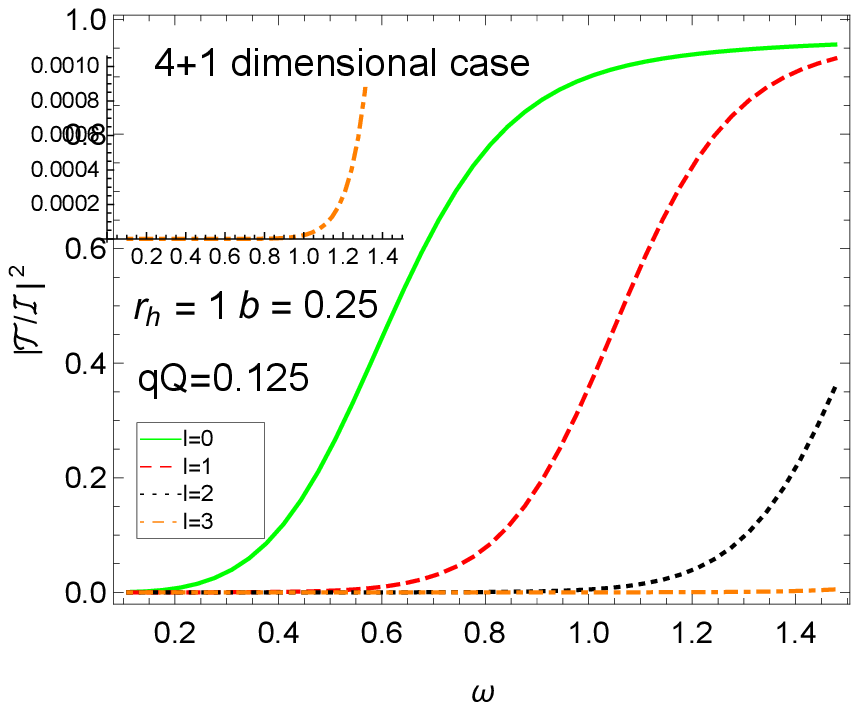}
\includegraphics[width=1\columnwidth]{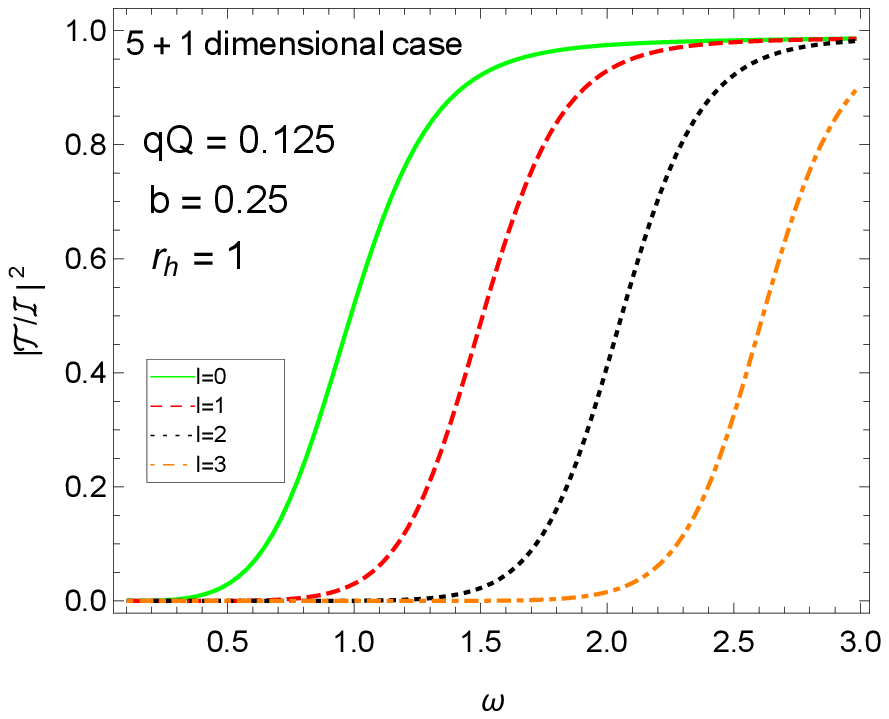}\includegraphics[width=1\columnwidth]{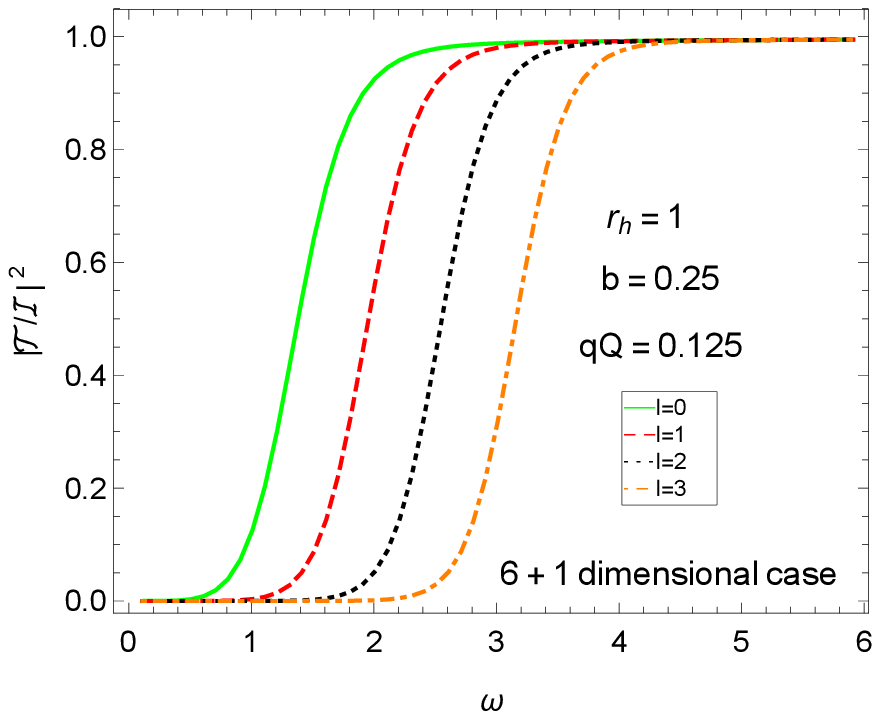}
\caption{
The same as Fig.~\ref{Figwb}, for different values of $l$.
The calculations are carried out with $r_h=1$, $qQ=\frac{1}{8}$, and $b=\frac{1}{4}$.}
\lb{Figwl}
\end{figure*}

\begin{figure*}[tbp]
\centering
\includegraphics[width=1\columnwidth]{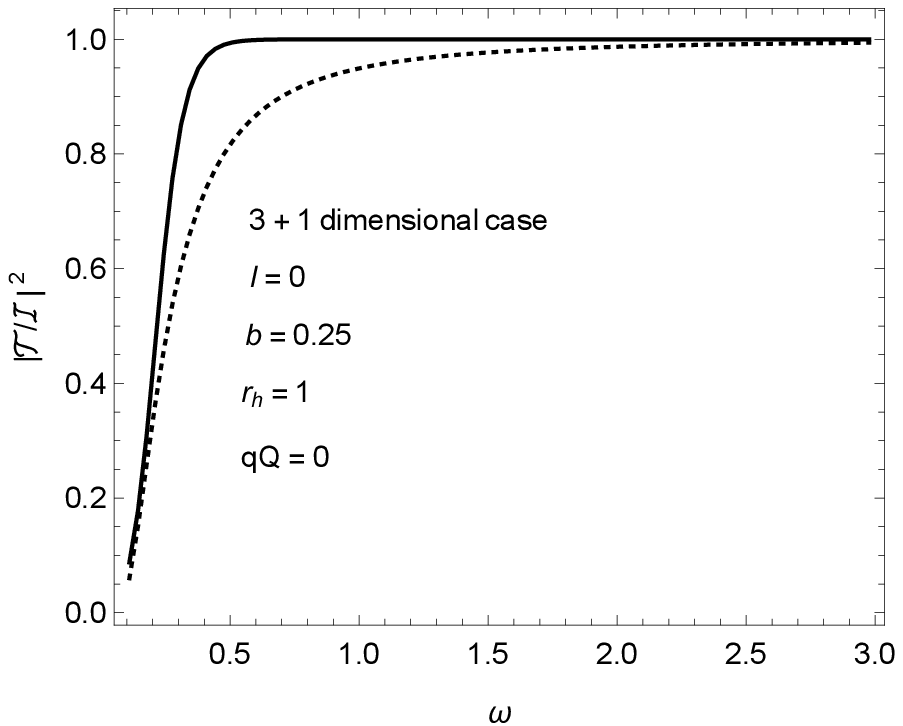}\includegraphics[width=1\columnwidth]{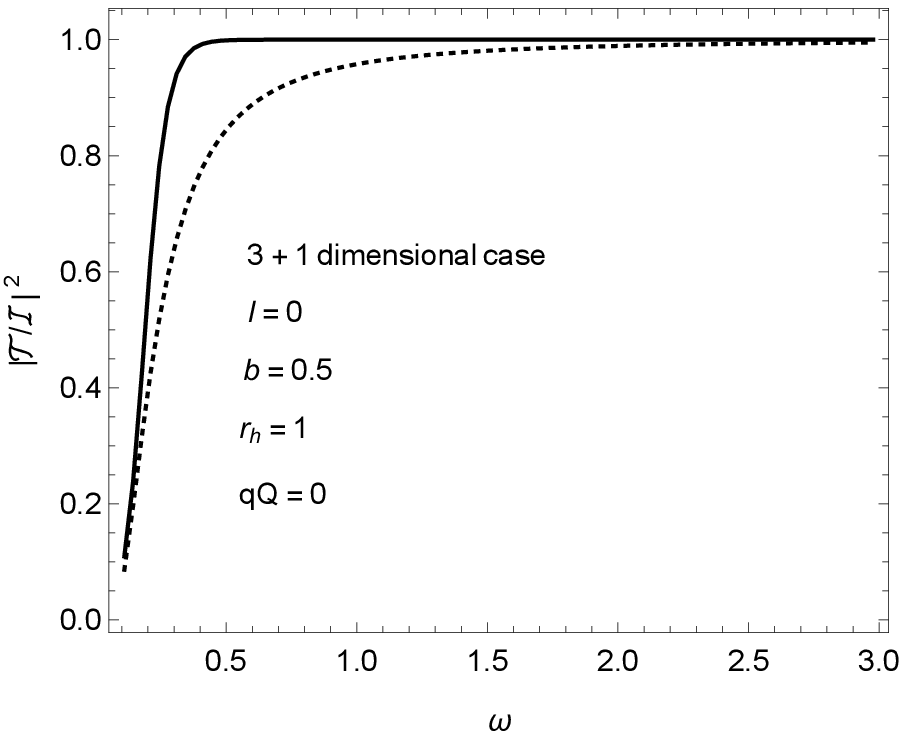}
\includegraphics[width=1\columnwidth]{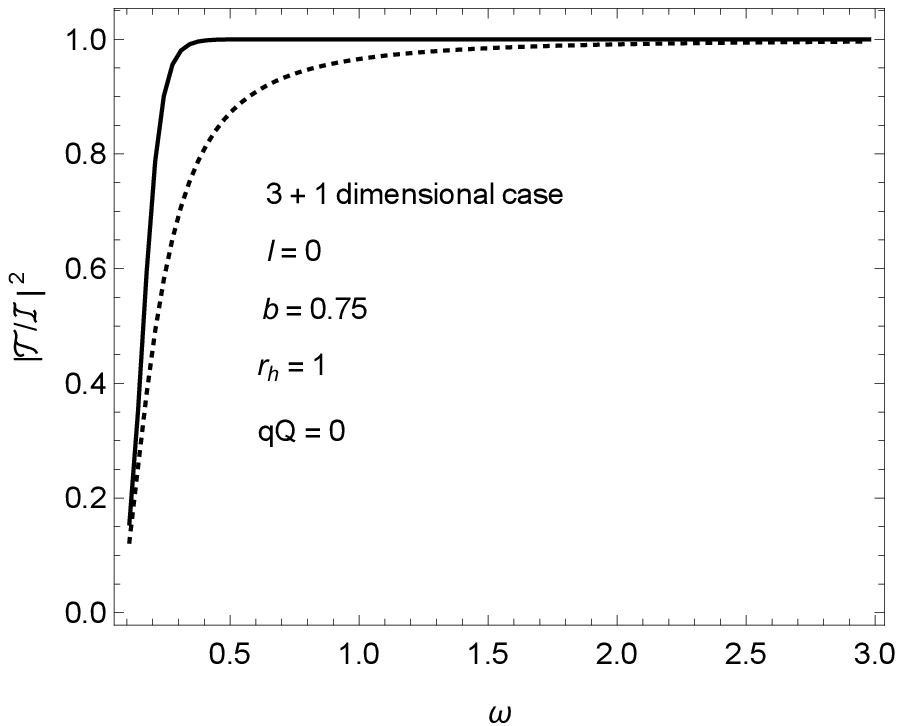}\includegraphics[width=1\columnwidth]{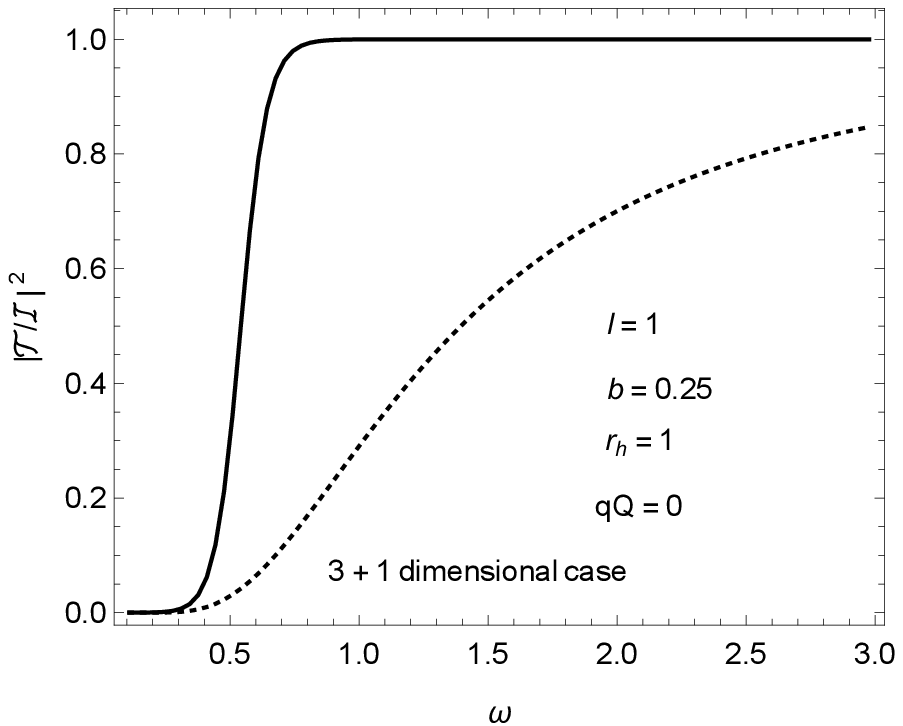}
\includegraphics[width=1\columnwidth]{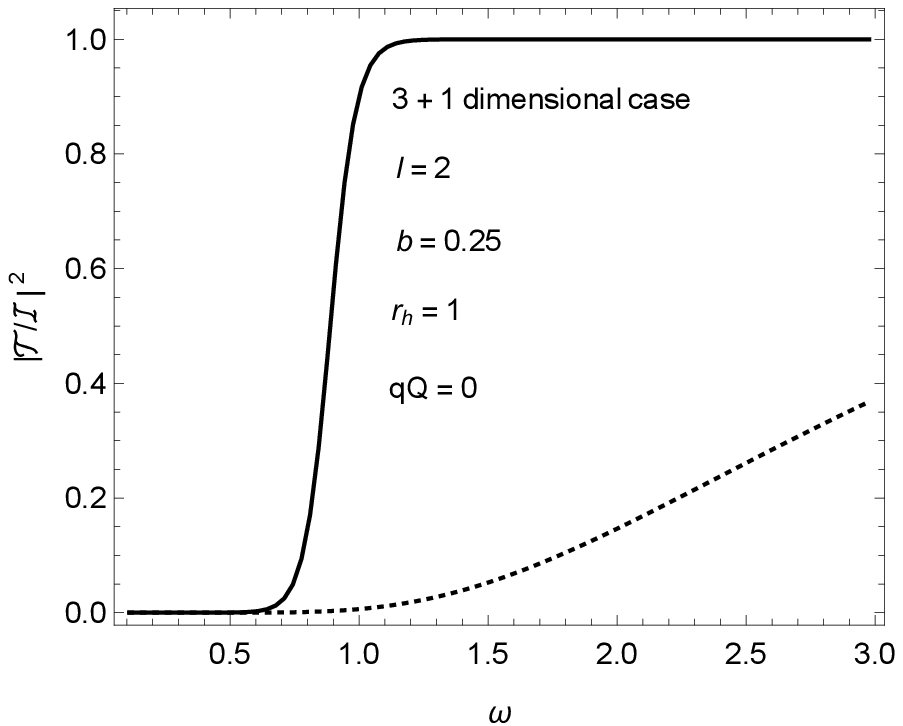}\includegraphics[width=1\columnwidth]{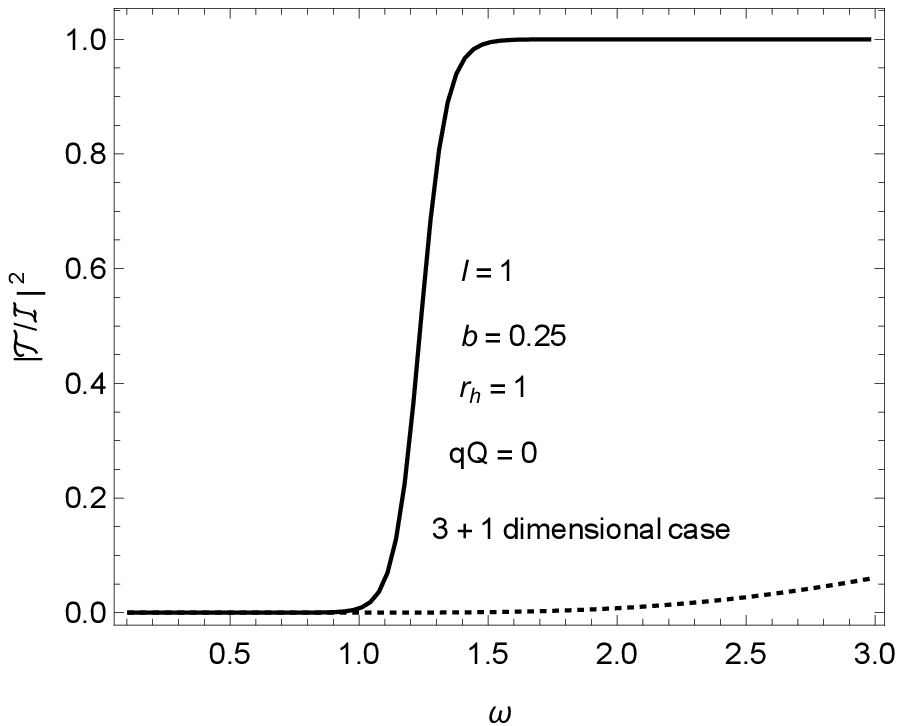}
\caption{
The same as Fig.~\ref{Figwb}, for different values of $b$ and $l$ in (3+1) dimensional spacetime.
The calculations are carried out with $r_h=1$ and $qQ=0$.
The calculated results (shown in solid curves) are compared with the low boundary (shown in dotted curves) obtained in Ref.~\cite{agr-bh-superradiance-07}.}
\lb{Figlimit}
\end{figure*}

\begin{figure*}[tbp]
\centering
\includegraphics[width=1\columnwidth]{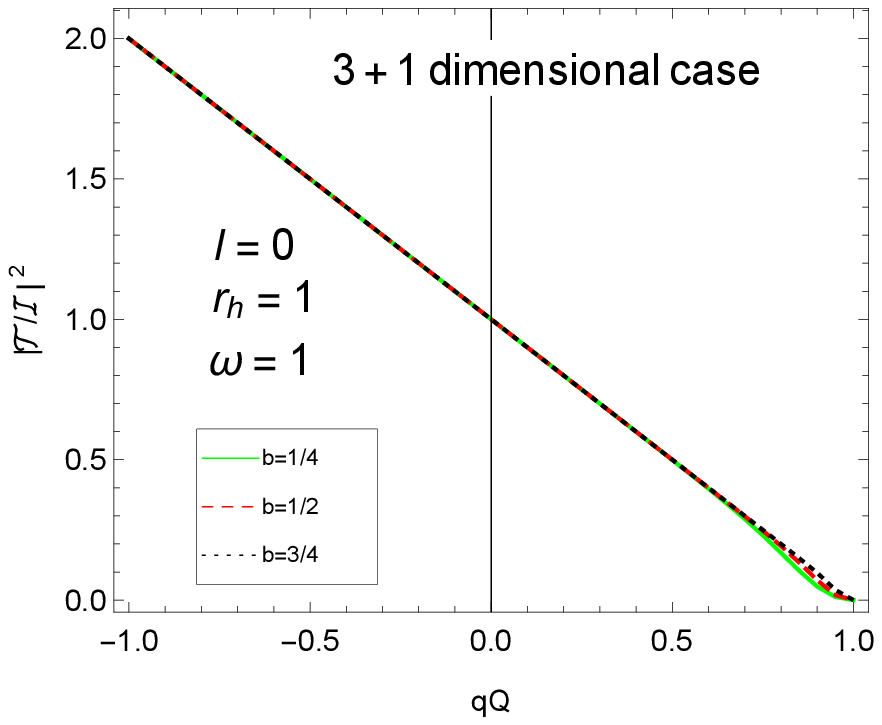}\includegraphics[width=1\columnwidth]{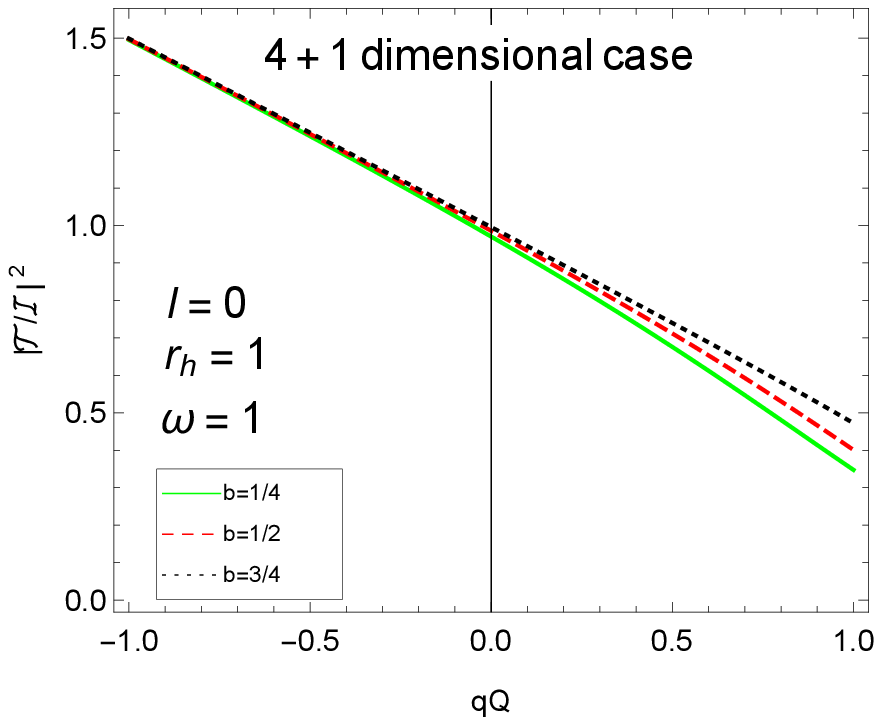}
\includegraphics[width=1\columnwidth]{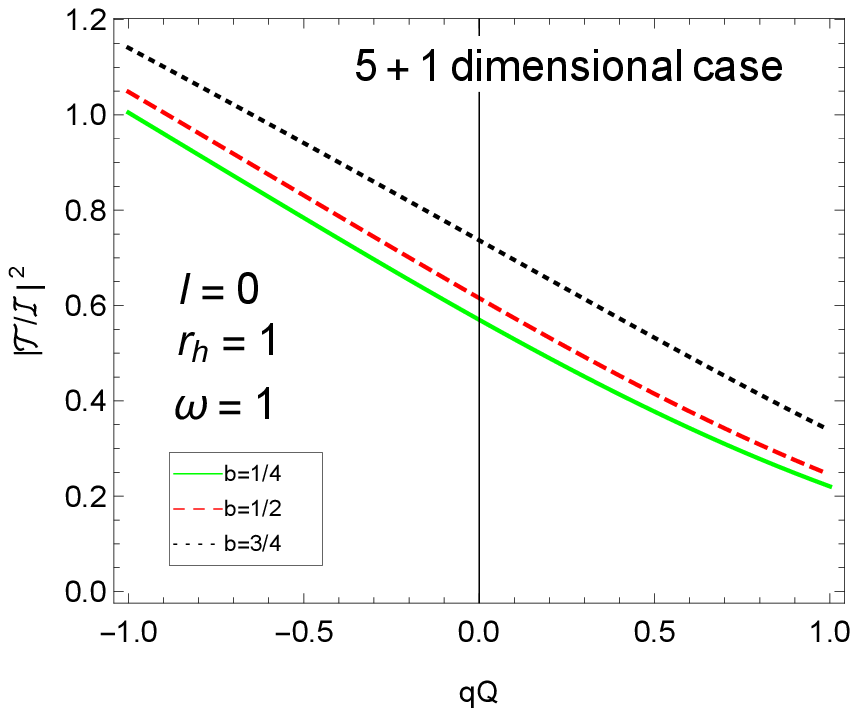}\includegraphics[width=1\columnwidth]{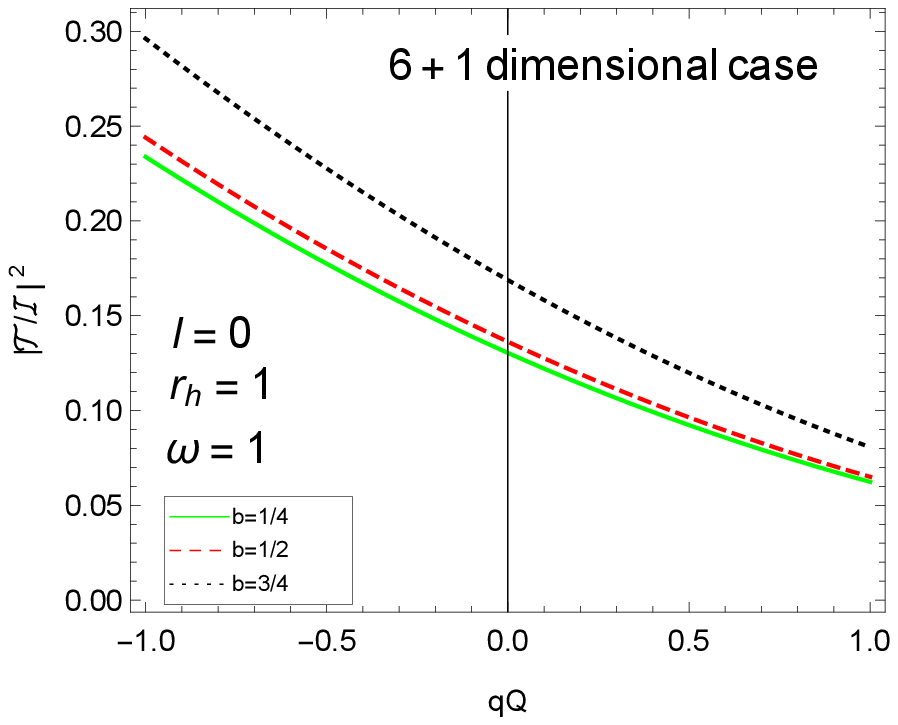}
\caption{
The calculated transmission coefficient $\left|\frac{\mathcal{T}}{\mathcal{I}}\right|^2$ for massless scsalar field as a function of $qQ$ for different values of $b$.
The calculations are carried out with $r_h=1$, $\omega=1$, and $l=0$.}
\lb{Figqb}
\end{figure*}

\begin{figure*}[tbp]
\centering
\includegraphics[width=1\columnwidth]{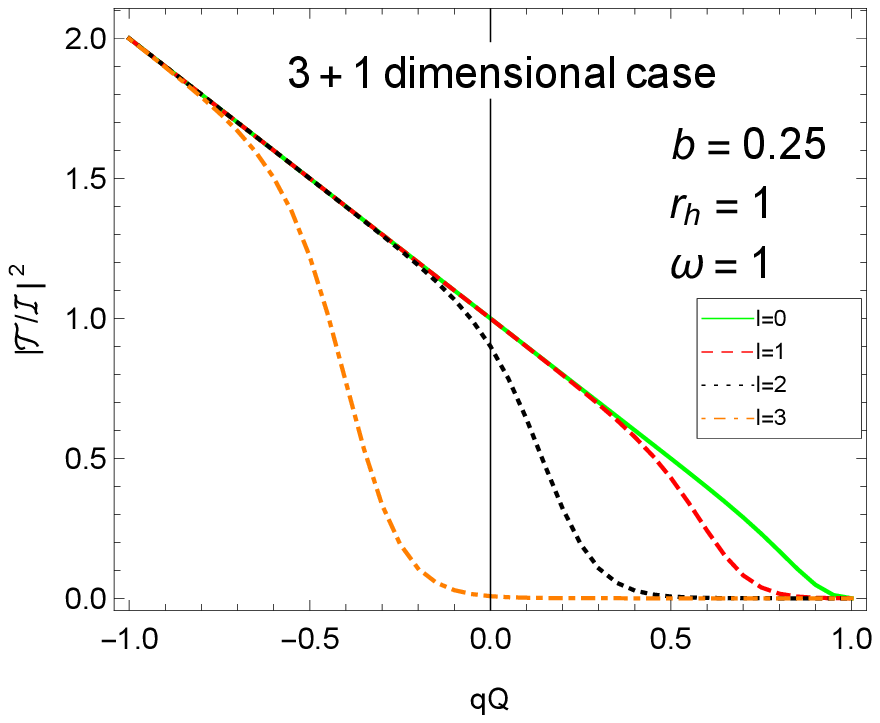}\includegraphics[width=1\columnwidth]{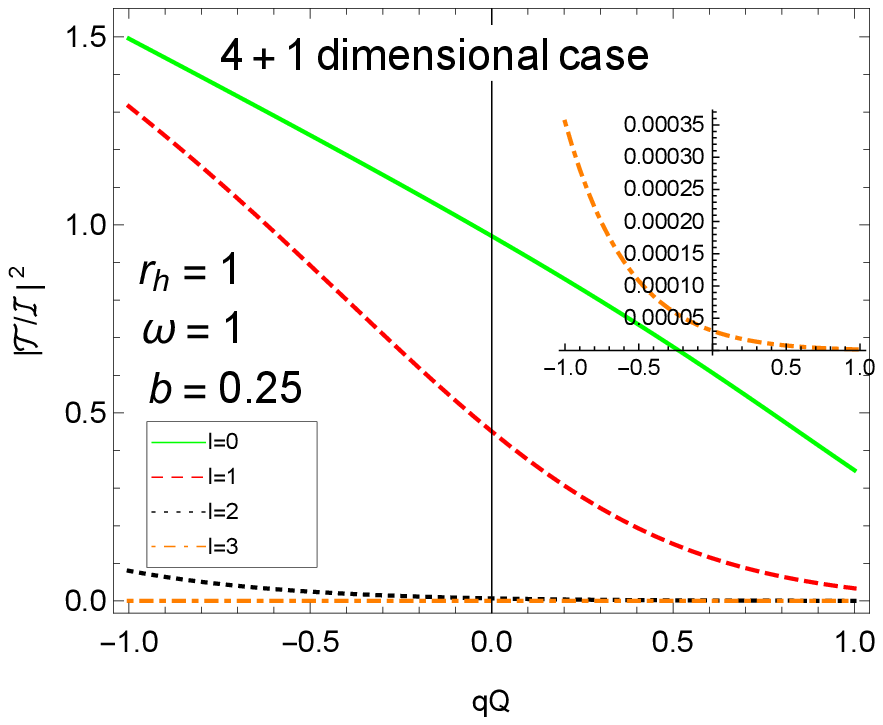}
\includegraphics[width=1\columnwidth]{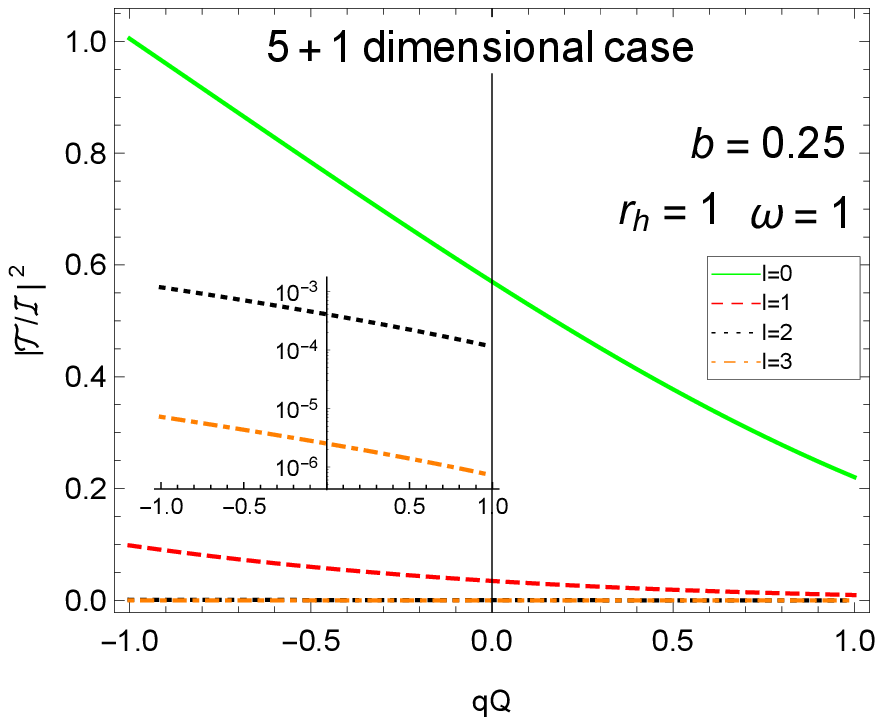}\includegraphics[width=1\columnwidth]{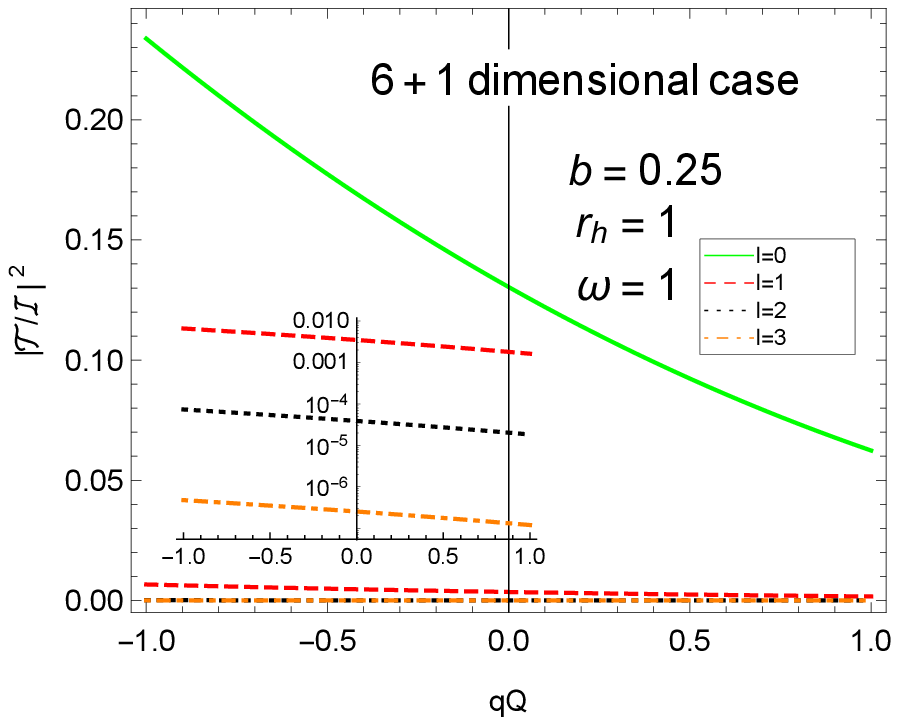}
\caption{
The calculated transmission coefficient $\left|\frac{\mathcal{T}}{\mathcal{I}}\right|^2$ for massless scsalar field as a function of $qQ$ for different values of $l$.
The calculations are carried out with $r_h=1$, $\omega=1$, and $b=\frac{1}{4}$.}
\lb{Figql}
\end{figure*}

%\begin{figure}[tbp]
%\centering
%\includegraphics[width=1\columnwidth]{FigC.eps}
%\caption{
%The calculated transmission coefficient $\left|\frac{\mathcal{T}}{\mathcal{I}}\right|^2$ as a function of the frequency $\omega$ for different types of fields.
%The calculations are carried out with $r_h=1$, $q=0$, $b=\frac{1}{4}$, and $l=2$.}
%\lb{FigC}
%\end{figure}

In Figs.~\ref{Figwb} -~\ref{Figlimit} we present the resultant transmission coefficient as a function of the frequcency, $\left|\frac{\mathcal{T}}{\mathcal{I}}\right|^2$ vs. $\omega$, for massless scsalar field.
The calculations are carried out by varying the parameter of the metric $b$ and the orbital quantum number $l$.
It is observed that the transmission coefficient monotonically increases with increasing frequency.
As shown in Fig.~\ref{Figweb}, they are in agreement with universal analytic results~\cite{agr-bh-superradiance-05,agr-bh-superradiance-06} in the low frequency region $\omega \ll T_h$ and $\omega r_h \ll 1$, as it rises from zero where the frequency vanishes. 
The physical interpretation is that at very low frequency, the size of the wavelength is much significant than that of the black hole.
Therefore an incident wave from infinity is virtually unaffected by the presence of the latter.
On the other hand, the transmission coefficient eventually saturates to some value at the limit of high frequency.
In Fig.~\ref{Figlimit}, the obtained results are also compared against the lower bound estimated in Ref.~\cite{agr-bh-superradiance-07}.
From Figs.~\ref{Figwb} and~\ref{Figwl}, at four dimension, it is found that resultant transmission coefficients all converge to the same curve, for given $r_h$ and $qQ$ and different values of $b$ and $l$.
However, as one goes to higher-dimensional cases, the convergence becomes more slowly.
Moreover, the difference between states with different $b$ and $l$ becomes more significant.

In Figs.~\ref{Figqb} and~\ref{Figql}, we investigate the transmission coefficient as a function of the product of the charge of the particle and the black hole $qQ$.
For all the cases, the transmission rate is found to be a monotonically decreasing function of $qQ$.
The dependence is mostly linear for given $b$ and for $l=0$.
For larger $l$, the curves show a twisted feature, and in the four-dimensional case, they are observed to converge together eventually.
In particular, it is shown that for $qQ<0$, as the magnitude becomes more significant, the transmission coefficient further increases and eventually exceeds one.
In other words, the resultant spectral flux is amplified by the effective potential.
This feature is reminiscent of the superradiance~\cite{agr-bh-superradiance-review-01}, which occurs when the frequency is less than a particular value related to the charge of the black hole.
By comparing the results for different spacetime dimensions, it is found that the difference between states with different $b$ and $l$ becomes more pronounced for higher dimensional spacetimes.
The slope of the monotonical dependence decreases as $l$ increases, and it becomes more evident as the dimension of spacetime increases. 
For example, for the cases of $5+1$ and $6+1$ metrics, the transmission coefficient of $l=0$ increases mostly linearly with decreasing $qQ$, apart from that the slope in $5+1$ is larger.
However, the corresponding slopes for the cases with $l=1, 2, 3$ are much less significant when compared to the former. 

%(maybe some figures can be merged.XXXX)
%(Is it possible to compare to the results of superradiance in RN metric?XXXX)

%In Fig.~\ref{FigC}, we show the calculated transmission coefficient as a function of the frequency, but for scalar, electromagnetic and gravitational fields with odd parity.
%A similar trend is demonstrated regarding their dependence on the frequency.
%Therefore, it is understood that the features found in Figs.~\ref{Figwb} and~\ref{Figwl} are robust as they remain unchanged for different types of fields.

\section{Concluding remarks}

To summarize, the Hawking radiation in Reissner-Nordstr\"om black hole spacetime regarding an observer resides at infinity is studied.
We evaluate the frequency-dependent transmission coefficient after discussing the thermal radiation emitted in the vicinity of the horizon.
It is found that the transmission coefficients approaches zero as the frequency of the emitted particle vanishes.
It is a monotonically increasing function of the frequency and saturates when the frequency is more significant.
In four-dimensional spacetime, this feature is shown to be mostly independent of the parameters, for given $r_h$ and $qQ$.
For higher-dimensional spacetimes, the difference between states with different $b$ and $l$ becomes more pronounced.
In particular, it is found that the transmission coefficient exceeds one when $qQ<0$ and the magnitude becomes large enough. 
This indicates that the spectral flux is further amplified during the course of traversing the curved spacetime.

\renewcommand{\theequation}{4.\arabic{equation}} \setcounter{equation}{0}

\begin{acknowledgments}
We gratefully acknowledge the financial support from
Funda\c{c}\~ao de Amparo \`a Pesquisa do Estado de S\~ao Paulo (FAPESP),
Funda\c{c}\~ao de Amparo \`a Pesquisa do Estado do Rio de Janeiro (FAPERJ),
Conselho Nacional de Desenvolvimento Cient\'{\i}fico e Tecnol\'ogico (CNPq),
Coordena\c{c}\~ao de Aperfei\c{c}oamento de Pessoal de N\'ivel Superior (CAPES),
and National Natural Science Foundation of China (NNSFC) under contract Nos. 11805166,11673008, and 11922303.
\end{acknowledgments}

\section*{Appendix: An adapted method for numerical integration}

This method was initially utilized for the calculation of superradiance~\cite{agr-bh-superradiance-01}.
For the present study, it can be readily adapted for the numerical calculations.
The code is implemented in terms of {\it Mathematica} notebook. 
For the purpose of the present study, we introduce the following adaptations.
First, we rewrite the radial equation by the coordinate transform $x=1-\frac{r_h}{r}$, so that $x=1$ as $r\rightarrow\infty$ and $x=0$ as $r=r_h$. 
By expanding the function at $x=0$ and $x=1$, one can numerically integrate the Schrodinger-type equation from the region near $x=0$ to that near $x=1$.
The code's efficiency lies in its significant accuracy of the numerical implementation for the integration in {\it Mathematica}.

\bibliographystyle{h-physrev}
\bibliography{references_qian}

\begin{thebibliography}{10}

\bibitem{agr-bh-thermodynamics-01}
J.~M. Bardeen, B.~Carter, and S.~Hawking,
\newblock Commun. Math. Phys. {\bf 31}, 161 (1973).

\bibitem{agr-bh-thermodynamics-02}
J.~D. Bekenstein,
\newblock Phys. Rev. D {\bf 7}, 2333 (1973).

\bibitem{agr-bh-thermodynamics-03}
J.~D. Bekenstein,
\newblock Phys. Rev. D {\bf 9}, 3292 (1974).

\bibitem{agr-bh-thermodynamics-04}
J.~D. Bekenstein,
\newblock Phys. Rev. {\bf D49}, 1912 (1994), gr-qc/9307035.

\bibitem{adscft-holo-01}
C.~R. Stephens, G.~'t~Hooft, and B.~F. Whiting,
\newblock Class. Quant. Grav. {\bf 11}, 621 (1994), gr-qc/9310006.

\bibitem{adscft-holo-03}
L.~Susskind,
\newblock J. Math. Phys. {\bf 36}, 6377 (1995), hep-th/9409089.

\bibitem{adscft-02}
E.~Witten,
\newblock Adv. Theor. Math. Phys. {\bf 2}, 253 (1998), hep-th/9802150.

\bibitem{adscft-01}
J.~M. Maldacena,
\newblock Int. J. Theor. Phys. {\bf 38}, 1113 (1999), hep-th/9711200,
\newblock [Adv. Theor. Math. Phys.2,231(1998)].

\bibitem{agr-hawking-radiation-01}
S.~W. Hawking,
\newblock Nature {\bf 248}, 30 (1974).

\bibitem{agr-hawking-radiation-02}
S.~W. Hawking,
\newblock Commun. Math. Phys. {\bf 43}, 199 (1975),
\newblock [Erratum: Commun. Math. Phys.46,206(1976)].

\bibitem{agr-hawking-radiation-06}
P.~Kraus and F.~Wilczek,
\newblock Nucl. Phys. {\bf B433}, 403 (1995), gr-qc/9408003.

\bibitem{agr-hawking-radiation-07}
M.~K. Parikh and F.~Wilczek,
\newblock Phys. Rev. Lett. {\bf 85}, 5042 (2000), hep-th/9907001.

\bibitem{agr-hawking-radiation-09}
Q.-Q. Jiang, S.-Q. Wu, and X.~Cai,
\newblock Phys. Rev. {\bf D73}, 064003 (2006), hep-th/0512351,
\newblock [Erratum: Phys. Rev.D73,069902(2006)].

\bibitem{agr-hawking-radiation-10}
J.~Zhang and Z.~Zhao,
\newblock Phys. Lett. {\bf B638}, 110 (2006), gr-qc/0512153.

\bibitem{agr-hawking-radiation-12}
K.~Srinivasan and T.~Padmanabhan,
\newblock Phys. Rev. {\bf D60}, 024007 (1999), gr-qc/9812028.

\bibitem{agr-hawking-radiation-13}
M.~Angheben, M.~Nadalini, L.~Vanzo, and S.~Zerbini,
\newblock JHEP {\bf 05}, 014 (2005), hep-th/0503081.

\bibitem{agr-hawking-radiation-14}
R.~Kerner and R.~B. Mann,
\newblock Phys. Rev. {\bf D73}, 104010 (2006), gr-qc/0603019.

\bibitem{agr-hawking-radiation-15}
R.~Li, J.-R. Ren, and S.-W. Wei,
\newblock Class. Quant. Grav. {\bf 25}, 125016 (2008), 0803.1410.

\bibitem{agr-hawking-radiation-16}
K.~Lin and S.-Z. Yang,
\newblock Phys. Rev. {\bf D79}, 064035 (2009).

\bibitem{agr-hawking-radiation-17}
K.~Lin and S.~Yang,
\newblock Phys. Lett. {\bf B674}, 127 (2009).

\bibitem{agr-hawking-radiation-18}
K.~Lin and S.-Z. Yang,
\newblock Chin. Phys. {\bf B20}, 110403 (2011).

\bibitem{agr-bh-superradiance-01}
D.~N. Page,
\newblock Phys. Rev. {\bf D13}, 198 (1976).

\bibitem{agr-bh-superradiance-06}
T.~Harmark, J.~Natario, and R.~Schiappa,
\newblock Adv. Theor. Math. Phys. {\bf 14}, 727 (2010), 0708.0017.

\bibitem{agr-bh-superradiance-03}
W.~G. Unruh,
\newblock Phys. Rev. {\bf D14}, 3251 (1976).

\bibitem{agr-qnm-05}
L.~Motl and A.~Neitzke,
\newblock Adv. Theor. Math. Phys. {\bf 7}, 307 (2003), hep-th/0301173.

\bibitem{agr-bh-superradiance-04}
A.~Neitzke,
\newblock (2003), hep-th/0304080.

\bibitem{agr-bh-superradiance-07}
T.~Ngampitipan and P.~Boonserm,
\newblock Int. J. Mod. Phys. {\bf D22}, 1350058 (2013), 1211.4070.

\bibitem{agr-bh-superradiance-05}
S.~R. Das, G.~W. Gibbons, and S.~D. Mathur,
\newblock Phys. Rev. Lett. {\bf 78}, 417 (1997), hep-th/9609052.

\bibitem{agr-bh-superradiance-review-01}
R.~Brito, V.~Cardoso, and P.~Pani,
\newblock Lect. Notes Phys. {\bf 906}, pp.1 (2015), 1501.06570.

\end{thebibliography}
\end{document}